# Molecular Theory for Self Assembling Mixtures of Patchy Colloids and Colloids with Spherically Symmetric Attractions: The Single Patch Case

Bennett D. Marshall and Walter G. Chapman.[1]
Department of Chemical and Biomolecular Engineering
Rice University
6100 S. Main
Houston, Texas 77005

## Abstract

In this work we develop the first theory to model self assembling mixtures of single patch colloids and colloids with spherically symmetric attractions. In the development of the theory we restrict the interactions such that there are short ranged attractions between patchy and spherically symmetric colloids, but patchy colloids do not attract patchy colloids and spherically symmetric colloids do not attract spherically symmetric colloids. This results in the temperature, density and composition dependent reversible self assembly of the mixture into colloidal star molecules. This type of mixture has been recently synthesized by grafting of complimentary single stranded DNA [Feng et al., Advanced Materials **25** (20), 2779-2783 (2013)]. As a quantitative test of the theory, we perform new monte carlo simulations to study the self assembly of these mixtures; theory and simulation are found to be in excellent agreement.

---

[1] Author to whom correspondence should be addressed.  Tel: (1) 713.348.4900.  Fax: (1) 713.348.5478.  Email: wgchap@rice.edu.

**I: Introduction**

The controlled, reversible and temperature dependant self assembly of anisotropic colloidal molecules will provide a new route to smart materials in the upcoming century.[1] A promising approach to controlled self assembly is the introduction of anisotropic potentials between spherical colloids by inclusion of some number of short ranged attractive surface patches.[2] These "patchy" colloids have been synthesized by glancing angle deposition[3,4], the polymer swelling method[5] and by stamping the colloids with patches of single stranded DNA.[6]

Theoretical modeling of patchy colloid fluids is complicated by the anisotropic nature of the intermolecular interactions as well patch saturation. Patch saturation is due to the fact that if two colloids share an attractive bond between two patches their hard cores may prevent, depending on patch size, a third colloid from approaching and sharing in the attraction bond. These are also two of the challenging features of developing a primitive model for hydrogen bonding.[7] A particularly successful approach for modeling hydrogen bonding fluids is the multi – density cluster theory of Wertheim[8-11] which explicitly accounts for the anisotropic nature of the hydrogen bonding potential, as well as bond saturation which he achieves through exact graph cancelation. Due to the similarities between patchy colloids and hydrogen bonding fluids, Wertheim's theory has been extensively applied to model the self assembly and phase equilibria of patchy colloid fluids.[12-23]

For single patch colloids (or many patch colloids in first order perturbation theory) Wertheim's theory takes on a very simple form if patch size is restricted such that each patch can bond at most once (single bonding condition), and contributions to the fundamental graph sum which contain more than one path of attraction bonds can be neglected (single chain approximation). The single chain approximation allows the perturbation theory to be written in terms of the hard sphere reference system correlation functions. Recently[24-28], there has been a significant amount of work going beyond the single bonding condition to allow multiple bonds per patch. To allow for $n$ bonds per patch the correct contributions to the free energy for $n + 1$ body association interactions between colloids must be included in the theory. This is only practical for small $n$ due to the rapidly increasing complexity as $n$ is increased. So far, these theoretical approaches have only accounted for the possibility of patches bonding twice ($n = 2$)[24-28].

In a recent paper[6], researchers synthesized mixtures of patchy $p$ and spherically symmetric $s$ colloids by binding DNA to the surfaces of the colloids. The $p$ colloids had a single sticky patch terminated with type $A$ single stranded DNA sticky ends and the $s$ colloids were uniformly coated with DNA terminated with complementary type $B$ single stranded DNA sticky ends. The DNA types were chosen such that there were $AB$ attractions but no $AA$ or $BB$ attractions. That is, $s$ colloids attract the anisotropic $p$ colloids, but $s$ colloids do not attract other $s$ colloids and $p$ colloids do not attract other $p$ colloids. It was shown that this mixture would reversibly self assemble into clusters where a single $s$ colloid would be bonded to some number of $p$ colloids into colloidal star molecules consisting of $n$ arms ($n$ patchy colloids). Of course $n$ is not uniform and falls onto some distribution.

In this paper we wish to derive a new theory to model this type of $s$ and $p$ colloid mixture. We will choose the patch size of the $p$ colloid such that the single bonding condition holds; however, since the $s$ colloid is a single uniform patch we must account for the fact that it can bond multiple times going well beyond the single bonding condition. Actually, the maximum number of bonds will be determined by the maximum number of $p$ colloids which can be fit into the "bonding shell" of the $s$ colloid. For equal sized $p$ and $s$ colloids this number must be a minimum of 12 which corresponds to the coordination number of hexagonal cubic closest packing of hard spheres. The fact that the *s* colloids do not attract each other will allow for the conceptually straight forward development of a theory which accurately describes the self assembly of these mixtures. We will develop the one patch theory in Wertheim's two density formalism[8, 9] for 1 site associating fluids. To account for the fact that the *s* colloids can bond to multiple *p* colloids we must include the contribution for each type of associated cluster in the free energy. To quantitatively test the new theory we perform new monte carlo simulations to test the effect of temperature, density and composition on the internal energy, pressure, average solvation number of *s* colloids, *n* distribution of s colloids and fraction of *p* colloids bonded. The theory and simulation are found to be in excellent agreement.

**II: Theory**

In this section we derive the theory for a two component mixture of patchy colloids, denoted *p*, and spherically symmetric colloids, denoted *s*. We consider the case that both colloids have the same diameter *d*. The *p* colloid has a single attractive type *A* patch the size of which is determined by the critical angle $\beta_c^{(p)}$ which defines the solid angle of the patch as $2\pi(1-\cos\beta_c^{(p)})$. The *s* colloid has a single large type *B* patch covering the entire surface of the colloid with a critical angle $\beta_c^{(s)} = 180°$. We allow *AB* attractions, but no *AA* or *BB* attractions. This model draws inspiration from recent experiments where researchers synthesized mixtures of *s* and *p* colloids by binding DNA to the surfaces of the colloids.[6] The *p* colloids had a patch of *A* type single stranded DNA and the *s* colloid had a complementary *B* type.[6] A diagram of these colloids can be found in Fig. 1.

Since there are no attractions between *p* colloids, their potential of interaction is simply that of a hard sphere system $\phi^{(p,p)}(r_{12}) = \phi_{HS}(r_{12})$. Similarly, since there are no attractions between *s* colloids their potential of interaction is also $\phi^{(s,s)}(r_{12}) = \phi_{HS}(r_{12})$. The potential of interaction between *s* and *p* colloids contains a hard sphere contribution and an attractive association contribution $\phi^{(s,p)}(r_{12}) = \phi_{HS}(r_{12}) + \phi_{as}^{(s,p)}(12)$. The attractions between *p* colloids[29] are commonly treated as conical square well association sites[30, 31]. Here we follow a similar approach with the *p* colloids treated as a hard sphere with a conical square well association site and the *s* colloid with a spherically symmetric square well association site giving the association potential

$$\phi_{as}^{(s,p)}(12) = \begin{cases} -\varepsilon_{AB}, & r_{12} \leq r_c \text{ and } \beta_A \leq \beta_c^{(p)} \\ 0 & \text{otherwise} \end{cases} \quad (1)$$

which states that if colloids 1 and 2 are within a distance $r_c$ of each other, and the $p$ colloid is oriented such that the angle between the site orientation vector and the vector connecting the two segments $\beta_A$ is less than the critical angle $\beta_c^{(p)}$, the two colloids are considered bonded and the energy of the system is decreased by a factor $\varepsilon_{AB}$. While simple in form, the parameters of this potential can be related back to the properties of the grafted single strand DNA. For instance, the critical radius $r_c$ will depend on the persistence length of the grafted DNA as well as solvent conditions, and the square well depth $\varepsilon_{AB}$ will depend on both DNA sequences and grafting densities among other things.[32]

We will develop the statistical mechanical theory to model the potential given in Eq. (1) in Wertheim's two density formalism[8,9] which gives the total density of species $k$ $\rho^{(k)}$ as the sum of the density of species $k$ bonded $\rho_b^{(k)}$ and the density of species $k$ not bonded $\rho_o^{(k)}$

$$\rho^{(k)} = \rho_o^{(k)} + \rho_b^{(k)} \qquad (2)$$

Using this definition of the densities, Wertheim developed an exact cluster expansion for associating 1 site molecules. In this formalism the free energy for a two component mixture can be written as

$$\frac{A - A_{HS}}{Vk_BT} = \sum_k \left( \rho^{(k)} \ln \frac{\rho_o^{(k)}}{\rho^{(k)}} - \rho_o^{(k)} + \rho^{(k)} \right) - \Delta c^{(o)}/V \qquad (3)$$

where $V$ is the system volume, $T$ is temperature and $A_{HS}$ is the free energy of the hard sphere reference system. The term $\Delta c^{(o)} = c^{(o)} - c_{HS}^{(o)}$ is the association contribution to the fundamental graph sum $c^{(o)}$ which encodes all intermolecular interactions. Wertheim's theory is typically applied within the single bonding condition, which states that each site can bond at most once. This is accurate for small to moderate patch size, but will incur error as patch size becomes large enough for multiple bonding of a patch to occur. For the current case the *s* colloid is a single spherical association site which can clearly not be modeled in the single bonding condition. The maximum number of bonds is simply the maximum number of *p* colloids $n^{max}$ which can pack in the *s* colloid's bonding shell $d \le r \le r_c$. To account for all association possibilities we will have to include contributions for each association possibility explicitly (one *s* colloid with one *p* colloid, two *p* colloids, three *p* colloids etc...). To accomplish this we decompose $\Delta c^{(o)}$ as

$$\Delta c^{(o)} = \sum_{n=1}^{n^{max}} \Delta c_n^{(o)} \tag{4}$$

where $\Delta c_n^{(o)}$ is the contribution for *n* patchy colloids bonded to a *s* colloid. Equation (4) is depicted in Fig. 2. We approximate $\Delta c_n^{(o)}$ in a generalization Wertheim's single chain approximation[10, 11] and consider all graphs consisting of a single associated cluster with *n* patchy colloids bonded to a *s* colloid with *n* association bonds. We also assume that the patch size on the *p* colloid is such that double bonding between one *p* colloid and multiple *s* colloids cannot occur. For each *n* this results in an infinite sum of graphs consisting of this associated cluster interacting with the hard sphere reference fluid which can be summed to yield

$$\Delta c_n^{(o)} = \frac{\rho_o^{(s)}\left(\rho_o^{(p)}\right)^n}{\widetilde{\Omega}^{n+1} n!} \int d(1)...d(n+1) \prod_{k=2}^{n+1} \left(f_{as}^{(s,p)}(1,k)\right) g_{HS}(1...n+1) \tag{5}$$

The notation $(1) = \{\vec{r}_1, \Omega_1\}$ where $\vec{r}_1$ is the position and $\Omega_1$ the orientation of colloid 1 in the cluster, $f_{as}^{(s,p)}(12) = \exp\left(-\phi_{as}^{(s,p)}(12)/k_B T\right) - 1$ is the association Mayer function, $\widetilde{\Omega} = 4\pi$ is the total number of orientations and $g_{HS}(1...n+1)$ is the $n+1$ body hard sphere reference correlation function. In Eq. (5) the $s$ colloid is labeled (1).

To evaluate the reference system correlation function $g_{HS}(1...n+1)$ we first use the definition of the hard sphere cavity correlation function

$$g_{HS}(1...n+1) = y_{HS}(1...n+1) \prod_{\substack{\text{all pairs} \\ \{l,k\}}} e_{HS}(r_{lk}) \tag{6}$$

Where $e_{HS}(r) = \exp\left(-\phi_{HS}(r)/k_B T\right)$ are the reference system $e$ bonds which serve to prevent hard sphere overlap in the cluster. We note from Fig. 2 that the associated clusters of $n$ patchy colloids are simply star molecules with $n$ arms of length 1. In Wertheim's perturbation theory, in the limit of infinitely strong association, the change in free energy in going from a mixture of hard spheres to a polyatomic star molecule consisting of $m$ hard spheres bonded at contact can be written as[33,34]

$$\frac{A_{star} - A_{HS}}{N_m k_B T} = -\ln y_{HS}^{(m)} \tag{7}$$

Where $y_{HS}^{(m)}$ is the $m$ body cavity correlation function averaged over the states of the star molecule and $N_m$ is the number of star molecules. While the correlation function $y_{HS}^{(m)}$ is not known in general we can approximate it as follows. Recently Marshall and Chapman[35] obtained an approximate general branched chain solution to the free energy (for spheres bonded at contact) in second order perturbation theory[11]. For star molecules with $n$ arms of length 1, $m = n + 1$, this solution is

$$\frac{A_{star} - A_{HS}}{N_m k_B T} = \begin{cases} -n \ln y_{HS}(d) + (n-3)\ln\left(\frac{1}{\sqrt{1+4\lambda}}\right) - 3\ln\left(\frac{1+\sqrt{1+4\lambda}}{2}\right) & \text{for } n > 1 \\ -\ln y_{HS}(d) & \text{for } n = 1 \end{cases} \quad (8)$$

Where $y_{HS}^{(2)}(d) = y_{HS}(d)$ is the pair cavity correlation function at hard sphere contact and[34]

$$\lambda = \frac{2}{3} \int_{\pi/3}^{\pi} \left(\frac{y_{HS}^{(3)}(d,d,2d\sin(\alpha/2))}{y_{HS}(d) y_{HS}(d)} - 1\right) \sin\alpha \, d\alpha \approx 0.2336\eta + 0.1067\eta^2 \quad (9)$$

In Eq. (9), $y_{HS}^{(3)}(d,d,2d\sin(\alpha/2))$ is the triplet cavity function where the first and third sphere are bonded at contact with the second and at an angle $\alpha$ to each other. Here $\eta$ is the packing fraction of the fluid.

For our current problem we will assume that the cavity function $y_{HS}(1...n+1)$ can be approximated by $y_{HS}^{(n+1)}$ which we obtain by solution of Eqns. (7) and (8) as

$$y_{HS}(1...n+1) \approx y_{HS}^{(n+1)} \approx y_{HS}(d)^n \delta^{(n)} \quad (10)$$

where

$$\delta^{(n)} = \begin{cases} (1+4\lambda)^{\frac{n-3}{2}} \left( \dfrac{1+\sqrt{1+4\lambda}}{2} \right)^3 & \text{for } n > 1 \\ 1 & \text{for } n = 1 \end{cases} \qquad (11)$$

Equations (10) and (11) constitute our approximation of the many body cavity correlation function.

With the association potential between $p$ colloids and $s$ colloids given by Eq. (1), Eq. (5) can now be simplified as

$$\Delta c_n^{(o)}/V = \frac{1}{n!} \rho_o^{(s)} \Delta^n \delta^{(n)} \Xi^{(n)} \qquad (12)$$

The term $\Delta$ is given by

$$\Delta = f_{as}^{(s,p)} \rho_o^{(p)} F y_{HS}(d) \qquad (13)$$

where $f_{as}^{(s,p)} = \exp(\varepsilon_{AB}/k_B T) - 1$ is the magnitude of the association Mayer function, $F = (1 - \cos\beta_c^{(p)})/2$ is the fractional patch coverage of the $p$ colloid. In Eq. (12) the integral $\Xi^{(n)}$ is given by

$$\Xi^{(n)} = \prod_{k=2}^{n+1} \left( \int_0^{2\pi} \int_{-1}^{1} \int_d^{r_c} d\phi_{1,k} \, d\cos\theta_{1,k} \, dr_{1,k} \, r_{1,k}^2 \right) \prod_{\substack{\text{all pairs} \\ \{l,k\}}} e_{HS}(r_{lk}) \qquad (14)$$

Equation (14) is simply a single cluster partition function for a cluster of 1 spherical colloid and $n$ patchy colloids. Here $\phi_{1,k}$, $\theta_{1,k}$ and $r_{1,k}$ define the azimuthal angle, polar angle and radial distance of $p$ colloid $k$ in a spherical coordinate system centered on the $s$ colloid 1. We discuss the evaluation of Eq. (14) in detail in the section III.

Now that $\Delta c^{(o)}$ has been completely specified we can minimize the free energy with respect to monomer densities to obtain the following mass action equations

$$\rho^{(s)} = \rho_o^{(s)} + \rho_o^{(s)} \sum_{n=1}^{n_{max}} \frac{1}{n!} \Delta^n \delta^{(n)} \Xi^{(n)} \tag{15}$$

$$\rho^{(p)} = \rho_o^{(p)} + \rho_o^{(s)} \sum_{n=1}^{n_{max}} \frac{n}{n!} \Delta^n \delta^{(n)} \Xi^{(n)} \tag{16}$$

Comparing to Eq. (2), we see that the second term on the right hand side of these two equations is the density of bonded colloids $\rho_b^{(k)}$. For the $s$ colloids it is convenient to introduce the densities $\rho_n^{(s)}$, which represent the density of *s*pherically symmetric colloids which are bonded to $n$ patchy colloids. By conservation these densities must satisfy the relation

$$\rho^{(s)} = \sum_{n=0}^{n_{max}} \rho_n^{(s)} \tag{17}$$

Comparing Eqns. (15) and (17) we can deduce the relation for $n > 0$

$$\rho_n^{(s)} = \frac{\rho_o^{(s)}}{n!} \Delta^n \delta^{(n)} \Xi^{(n)} \qquad n > 0 \tag{18}$$

Introducing the fractions $X_n^{(k)} = \rho_n^{(k)} / \rho^{(k)}$ we can write the average cluster size $\bar{n}$ (average number of $p$ colloids bonded to an $s$ colloid) as

$$\bar{n} = \sum_{n=0}^{n^{max}} n X_n^{(s)} \tag{19}$$

Equations (15) and (16) are now simplified as

$$\frac{1}{X_o^{(s)}} = 1 + \sum_{n=1}^{n^{max}} \frac{1}{n!} \Delta^n \delta^{(n)} \Xi^{(n)} \tag{20}$$

and

$$X_o^{(p)} = 1 - \frac{x^{(s)}}{\left(1-x^{(s)}\right)} \frac{\sum_{n=1}^{n^{max}} \frac{n}{n!} \Delta^n \delta^{(n)} \Xi^{(n)}}{1 + \sum_{n=1}^{n^{max}} \frac{1}{n!} \Delta^n \delta^{(n)} \Xi^{(n)}} \tag{21}$$

where $x^{(s)}$ is the mole fraction of $s$ colloids. To obtain the monomer fractions, Eq. (21) is first solved numerically for $X_o^{(p)}$ and then Eq. (20) can be used to evaluate $X_o^{(s)}$.

Now we combine Eqns. (3) - (4), (12), (17) - (18) to simplify the free energy to the following form

$$\frac{A - A_{HS}}{Nk_B T} = x^{(s)} \ln X_o^{(s)} + \left(1 - x^{(s)}\right)\left(\ln X_o^{(p)} - X_o^{(p)} + 1\right) \tag{22}$$

where $N$ is the total number of colloids. Equations (20) – (22) give the fundamental equations for the theory of single patch colloids interacting with spherically symmetric colloids. The chemical potential, pressure and internal energy are all calculated in the appendix. In section III we discuss the evaluation of the cluster partition functions $\Xi^{(n)}$

**III: Evaluation of $\Xi^{(n)}$**

In this section we evaluate the cluster partition functions $\Xi^{(n)}$ given by Eq. (14). For $n = 1$ we solve the integral analytically to obtain

$$\Xi^{(1)} = \frac{4}{3}\pi\left(r_c^3 - d^3\right) = v_b \tag{23}$$

where $v_b$ is the bonding volume. We can also obtain an analytical solution for $n = 2$

$$\Xi^{(2)} = \frac{v_b^2}{2} + \pi^2\left(r_c^3 - d^2 r_c\right)^2 \tag{24}$$

The remaining $\Xi^{(n)}$ are evaluated numerically using monte carlo integration[29] as

$$\Xi^{(n)} = v_b^n P^{(n)} \tag{25}$$

The term $P^{(n)}$ is the probability that if we randomly generate $n$ patchy colloids in the bonding shell of the $s$ colloid that there is no hard sphere overlap. For the case $r_c = 1.1d$, Eq. (25) was evaluated using this Monte Carlo integration routine for cluster sizes $1 \leq n \leq 9$. However, for $n \geq 9$ this method becomes very inefficient due to the low probability of generating this many hard spheres in the bonding shell of the $s$ colloid and there being no overlap.

A much more efficient method to evaluate $P^{(n)}$ in this case is the following

$$P^{(n)} = P^{(n)}_{insert} P^{(n-1)} \tag{26}$$

Which states that the probability that $n$ randomly generated $p$ colloids will not have any hard sphere overlap, is simply the probability that $n - 1$ randomly generated $p$ colloids do not overlap multiplied by an insertion probability $P_{insert}^{(n)}$. This insertion probability is simply the probability that a randomly generated $p$ colloid in the bonding shell of the $s$ colloid, with $n - 1$ non-overlapping $p$ colloids already in place, will not overlap with any of the existing $n - 1$ patchy colloids. Mathematically the insertion probability is given by

$$P_{insert}^{(n)} = \left\langle \prod_{j=1}^{n-1} e_{HS}(r_{j,n}) \right\rangle_{n-1} \tag{27}$$

where $\langle \ \rangle_{n-1}$ represents an ensemble average in a cluster of $n - 1$ non-overlapping $p$ colloids in the bonding shell of a $s$ colloid. This is similar to Widom test particle insertions.[36] Solving Eq. (26) recursively we obtain

$$P^{(n)} = \prod_{k=1}^{n} P_{insert}^{(k)} \tag{28}$$

Equation (27) was evaluated using standard monte carlo simulation techniques.[29] In total $10^8 - 10^9$ configurations were generated with $10^8 - 10^9$ insertions used to evaluate the average in Eq. (27). Numerical calculations for the probabilities $P_{insert}^{(n)}$ and $P^{(n)}$ can be found in Table 1 for the case $r_c = 1.1d$. As expected, increasing $n$ results in a decrease in both $P_{insert}^{(n)}$ and $P^{(n)}$. For this case, the maximum number of $p$ colloids for which we obtained a non-zero insertion probability was found to be $n^{max} = 13$. This had a much lower probability than the case $n = 12$, which corresponds to hexagonal cubic closest packing if the colloids were restricted to bond at contact.

**IV: Simulations**

As a test of the theory, we perform new monte carlo simulations (not to be confused with the simulations discussed in section III) for the case of a mixture of *s* and *p* type colloids which are hard spheres with an additional attractive potential given by Eq. (1). Unless otherwise stated, we use the potential parameters $r_c = 1.1d$ and $\beta_c^{(p)} = 27°$ such that only single bonding of a *p* colloid will occur. Constant *NVT* (number of colloids, volume, temperature) simulations were performed using standard methodology.[29] Each *NVT* simulation was allowed to equilibrate for $10^8 - 10^9$ trial moves and averages where taken for another $10^8 - 10^9$ trial moves. A trial move consists of an attempted relocation of a *s* colloid or an attempted relocation and reorientation of a *p* colloid. For each simulation we used a total of *N* = 864 colloids. Constant *NPT* (number of colloids, pressure, temperature) simulations were performed in the same manner as the *NVT* simulations with the addition of an attempted volume change each *N* trial moves.

## V: Results

Now we use the theory derived in sections II and III to study the self assembly of mixtures of $s$ and $p$ colloids. At each point we will also compare to monte carlo simulation results to validate the new theory. We begin with a discussion of the dependence on $s$ colloid mole fraction $x^{(s)}$ when association energy (inverse temperature) $\varepsilon^* = 1/T^* = \varepsilon_{AB}/k_B T$ and density $\rho^* = \rho d^3$ are both held constant. Comparison of theory and simulation predictions of the excess internal energy $E^* = E^{AS}/Nk_B T$, average number of bonds per $s$ type colloid $\bar{n}$ and fraction of patchy colloids bonded $X_1^{(p)} = \rho_b^{(p)}/\rho^{(p)}$ can be found in Fig. 3, for both low $\rho^* = 0.2$ and high $\rho^* = 0.7$ density cases. Here the association energy is set at $\varepsilon^* = 7$.

For each case, $\bar{n}$ increases with decreasing $x^{(s)}$ reaching a maximum when $x^{(s)} \to 0$. The reasoning behind this is simple, when $x^{(s)}$ is small there is an abundance of $p$ colloids available to "solvate" the $s$ colloids. As $x^{(s)}$ is increased, $\bar{n}$ decreases because there are less $p$ colloids available for association due to a decreased fraction of $p$ colloids and competition with other $s$ colloids. Increasing density increases $\bar{n}$ as expected. The theory is in excellent agreement with simulation for $\rho^* = 0.2$ over the full range of $x^{(s)}$ and is in good agreement for the higher density case $\rho^* = 0.7$ over much of the $x^{(s)}$ range; however, at this higher density the accuracy of the theory decreases somewhat for $x^{(s)} \to 0$. It is in this limit that $\bar{n}$ is a maximum and the high order contributions $\Delta c_7^{(o)}$, $\Delta c_8^{(o)}$ etc… come into play. To evaluate these graphs we approximated the many body correlation functions $y_{HS}(1....n+1)$ by the approximation given by Eq. (10). We expect this to be most accurate at low densities and less accurate at higher densities. For this reason the accuracy of the theory decreases for $x^{(s)} \to 0$ and high density. That said, the overall agreement between theory and simulation is very good.

The fraction $X_1^{(p)}$ shows the opposite $x^{(s)}$ dependence as compared to $\bar{n}$. $X_1^{(p)}$ is a maximum for $x^{(s)} \to 1$ when there are an abundance of $s$ colloids available for association and a minimum for $x^{(s)} \to 0$ when there are few $s$ colloids available for association. The $x^{(s)}$ dependence of $E^*$ is more interesting. For small $x^{(s)}$, increasing the fraction of $s$ colloids increases association due to the fact that the system is $s$ colloid limited. This results in a more negative $E^*$. For large $x^{(s)}$, increasing the fraction of $s$ colloids decreases association due to the fact that the system is now $p$ colloid limited. This results in a less negative $E^*$. Near $x^{(s)} \sim 0.2$ association is a maximum resulting in a clear minimum in $E^*$ at both densities. We see that the minimum in $E^*$ shifts to lower $x^{(s)}$ as density is increased. This is due to the increase in $\bar{n}$ as density is increased. Theory and simulation are in excellent agreement.

The left panel of Fig. 4 shows calculations for the compressibility factor $Z$ at the same conditions shown in Fig. 3. At $x^{(s)} = 0$ there is no association in the system and $Z$ is that of a hard sphere fluid. Increasing $x^{(s)}$, when $x^{(s)}$ is small, results in a decrease in $Z$ as $s$ and $p$ type colloids associate into larger clusters. The opposite is true for large $x^{(s)}$ where increasing $x^{(s)}$ decreases association and results in an increase in $Z$. Like the internal energy $E^*$, $Z$ shows a distinct minimum due to these competing effects near $x^{(s)} \sim 0.2$ when association is maximized in the system. In the right panel of Fig. 4 we compare theory to *NPT* simulations at an association energy $\varepsilon^* = 7$ and mole fractions $x^{(s)} = 0$, 0.0579 and 0.174. The agreement between theory and simulation is good, although the theory slightly overpredicts $Z$ at high density for the case $x^{(s)} = 0.0579$. This is due to the underprediction of $\bar{n}$ in this regime, Fig. 3.

Now we will specifically consider the effect of association energy (inverse temperature) at constant density and composition. These results can be found in Figs. 5 and 6 for $\rho^* = 0.2$ and $\rho^* = 0.7$ respectively. At each density we perform calculations for $x^{(s)} = 0.5$ (case I), 0.174 (case II) and 0.0579 (case III). We begin our discussion of these figures with case I. For case I there is an equal number of $s$ and $p$ colloids, so we should expect the relation $\bar{n} = X_1^{(p)}$ to hold exactly for each $\varepsilon^*$; this is observed in both theory and simulation. When there is an abundance of $s$ colloids, as in case I, the entropic penalty of association is quite low. For this reason we see a rapid increase in $X_1^{(p)}$ even at low $\varepsilon^*$. This results in case I having the lowest $E^*$ for the given $x^{(s)}$ when $\varepsilon^*$ is low. Of course, for this case, when $\varepsilon^*$ is high $\bar{n}$ must reach a limiting value of 1 due to simple stoichiometry. Decreasing the mole fraction to $x^{(s)} = 0.174$ (case II) the fraction $X_1^{(p)}$ now increases more slowly with $\varepsilon^*$ due to the fact that there are less $s$ colloids available for association as compared to case I. For large $\varepsilon^*$, $\bar{n}$ reaches a limiting value of 4.76 which is simply the ratio of $p$ colloids to $s$ colloids. Of the three cases, case II has the most negative $E^*$ for strong association energies. In case III the ratio of $p$ to $s$ colloids is 814/50 = 16.28 which is greater than $n^{max}$. This means that there is no stoichiometric limit to $\bar{n}$. For this reason we see that $\bar{n}$ continues to increase over the studied range of $\varepsilon^*$. Also, $X_1^{(p)}$ does not approach 1 as in cases I and II. Comparing the low and high density cases (Figs. 5 and 6) we see that the results are qualitatively similar with association being enhanced for $\rho^* = 0.7$. The theory and simulation are in good agreement; however, for case III at $\rho^* = 0.7$ there is some error for the same reasons discussed previously.

Until now, for convenience, we have only considered the average number of bonds per $s$ colloid (solvation number) $\bar{n}$. However, the mole fraction of $s$ colloids in each cluster of $n$ patchy colloids $X_n^{(k)}$ is easily calculated through Eq. (18). We show these fractions for the cases

I – III, discussed above, in Fig. 7 at a density of $\rho^* = 0.2$ and association energies $\varepsilon^* = 6$ and 12. Simulations were performed for $\varepsilon^* = 6$ and are represented by the star symbols. First we will focus on the bottom panel, case III. As discussed above, for this case there are enough *p* colloids for the *s* colloids to become fully bonded. At an association energy of $\varepsilon^* = 6$, the average number of bonds per *s* colloid is $\bar{n} = 2.89$, and the distribution of $X_n^{(s)}$ is more or less Gaussian with significant contributions ranging between *n* = 0 and *n* = 6. Increasing the association energy to $\varepsilon^* = 12$ the shape of the distribution is similar to the $\varepsilon^* = 6$ case with $\bar{n}$ shifted to 9.32 with non-negligible contributions as high as *n* = 12. For case II, with a mole fraction $x^{(s)} = 0.174$, the $\bar{n}$'s shift to lower *n*, $\bar{n} = 2.14$ for $\varepsilon^* = 6$ and $\bar{n} = 4.72$ for $\varepsilon^* = 12$, due to the fact that the *p* colloids are now limiting with a maximum possible average solvation number of 4.76. Even with the low $\bar{n} = 2.14$ for $\varepsilon^* = 6$ there are still significant contributions for clusters with *n* = 5 patchy colloids. Lastly, for case I, there are an equal number of *p* and *s* colloids rendering a maximum average solvation number of 1. For this case the distributions are asymmetric with the majority of *s* colloids being in clusters $0 \leq n \leq 3$ with clusters of larger *n* contributing to a lesser degree. Theory and simulation are in excellent agreement for each case.

Lastly we will consider the effect of fractional patch coverage *F* of the patchy colloid on $\bar{n}$. In general, one has to account for the possibility of multiple bonds per patch once patch size increases beyond a certain point, about $\beta_c^{(p)} \sim 30°$ for $r_c = 1.1d$.[24, 27] For the current case, in which the patchy colloids do not self attract, the possibility of a patchy colloid with large patch size forming a double bond will vanish for small enough $x^{(s)}$. Simply put, if the *s* type colloids are very dilute the probability of a *p* colloid simultaneously interacting with more than one *s* colloid (regardless of patch size) is vanishing. In this regime the theory derived in this work should be accurate for the full range of patch sizes. We validate this in Fig. 8 which shows the $F$ dependence of $\bar{n}$ for the case $\varepsilon^* = 8$ and $\rho^* = 0.2$ at a spherical colloid mole fraction of $x^{(s)} = 0.00579$. As can be seen, $\bar{n}$ increases logarithmically with $F$. The theory and simulation are in excellent agreement. This logarithmic dependence can be explained by the following simple model. We can write the change in Helmholtz free energy due to forming a single bond as $\Delta A^b = \Delta U^b - T\left(\Delta S^b_{config} + \Delta S^b_{orient}\right)$, where $\Delta U^b = -\varepsilon_{AB}$ is the change in internal energy, $\Delta S^b_{config}$ is the change in configurational entropy and $\Delta S^b_{orient} \sim \ln F$ is the change in orientational entropy[6] due to bond formation. Since $\Delta U^b$ and $\Delta S^b_{config}$ are both independent of $F$ we can say

$$\frac{\partial \Delta A^b}{\partial F} \sim -\frac{1}{F} \qquad (29)$$

Equation (29) states that a small change in patch size results in a large decrease in the free energy for small $F$, while for larger $F$ the decrease in free energy upon increasing patch size is less. This is due to the fact that the penalty in decreased orientational entropy due to association is much less for large patches than for small patch sizes. This is the genesis of the logarithmic dependence observed in Fig. 8.

## VI: Conclusions

We have derived the first theory capable of modeling mixtures of patchy *p* and spherically symmetric *s* colloids. In the current derivation we have assumed that the *p* colloids have a single patch which can engage in a single bond to a *s* colloid only. The *s* colloids consist of a single spherically symmetric site which can bond to as many *p* colloids as can physically fit in the *s* colloids bonding shell with no overlap. We have enforced the restriction that the spherically symmetric colloid cannot bond to other spherically symmetric colloids. Inspiration for this model was drawn from the recent work of Feng et *al.*[6] who synthesized mixtures of spherical and patchy single stranded DNA coated nanoparticles which have the same type of restrictions as discussed above. The new theory was extensively tested against new monte carlo simulation results and was found to be accurate.

There are a number of ways to improve on the current model. The first is to extend the theory such that the patchy colloids can have multiple patches and the second is to allow the *s* colloids to bond to other *s* colloids. Extending the theory to include multiple patches on the *p* colloids is a complicated, but doable problem. Incorporating bonding between the spherically symmetric colloids is a much more daunting task. A more appropriate formalism to include this effect may be the cluster theory of Kalyuzhnyi and Stell[37] which is formulated to include association type interactions between molecules with spherically symmetric potentials. We will consider these extensions in a future paper.

**Appendix: Calculation of thermodynamic quantities**

In this appendix we calculate the chemical potential $\mu$, pressure $P$ and excess internal energy $E^{AS}$. The simplest way to calculate the chemical potential is to use the Euler – Lagrange equation supplied by Wertheim[8]

$$\frac{\mu^{(k)}}{k_B T} = \frac{\mu_{HS}^{(k)}}{k_B T} + \ln X_o^{(k)} - \frac{\partial \Delta c^{(o)}/V}{\partial \rho^{(k)}} \qquad (A1)$$

where $\mu_{HS}^{(k)}$ is the hard sphere reference chemical potential for component $k$. From Eq. (A1) we obtain for $k = \{s, p\}$

$$\frac{\mu^{(k)}}{k_B T} = \frac{\mu_{HS}^{(k)}}{k_B T} + \ln X_o^{(k)} - \bar{n}\rho^{(s)} \frac{\partial \ln y_{HS}(d)}{\partial \rho^{(k)}} - \sum_{n=1}^{n^{max}} X_n^{(s)} \rho^{(s)} \frac{\partial \ln \delta^{(n)}}{\partial \rho^{(k)}} \qquad (A2)$$

where $\bar{n}$ is the average number of bonds per $s$ type colloid and is given by Eq. (19). With the chemical potentials known the pressure is easily calculated through the relation

$$P = \sum_k \mu^{(k)} \rho^{(k)} - A/V \qquad (A3)$$

Lastly we obtain the excess internal energy as

$$\frac{E^{AS}}{N} = \frac{\partial}{\partial \beta}\left(\frac{\beta(A - A_{HS})}{N}\right) = x^{(s)} \frac{1}{X_o^{(s)}} \frac{\partial X_o^{(s)}}{\partial \beta} + (1 - x^{(s)}) \frac{\partial X_o^{(p)}}{\partial \beta}\left(\frac{1}{X_o^{(p)}} - 1\right) \qquad (A4)$$

**Tables:**

| $n$ | $P^{(n)}_{insert}$ | $P^{(n)}$ |
|---|---|---|
| 1 | 1 | 1 |
| 2 | 0.774 | 0.774 |
| 3 | 0.573 | 0.444 |
| 4 | 0.401 | 0.178 |
| 5 | 0.261 | 0.0463 |
| 6 | 0.153 | $7.11 \times 10^{-3}$ |
| 7 | 0.0790 | $5.61 \times 10^{-4}$ |
| 8 | 0.0340 | $1.91 \times 10^{-5}$ |
| 9 | 0.0111 | $2.12 \times 10^{-7}$ |
| 10 | $6.97 \times 10^{-3}$ | $1.48 \times 10^{-9}$ |
| 11 | $3.37 \times 10^{-3}$ | $4.98 \times 10^{-12}$ |
| 12 | $2.23 \times 10^{-4}$ | $1.11 \times 10^{-15}$ |
| 13 | $\sim 10^{-9}$ | $\sim 10^{-24}$ |

**Table 1:** Insertion probabilities $P^{(n)}_{insert}$ (27) and generation probabilities $P^{(n)}$ (26) calculated for $r_c = 1.1d$

**Figures:**

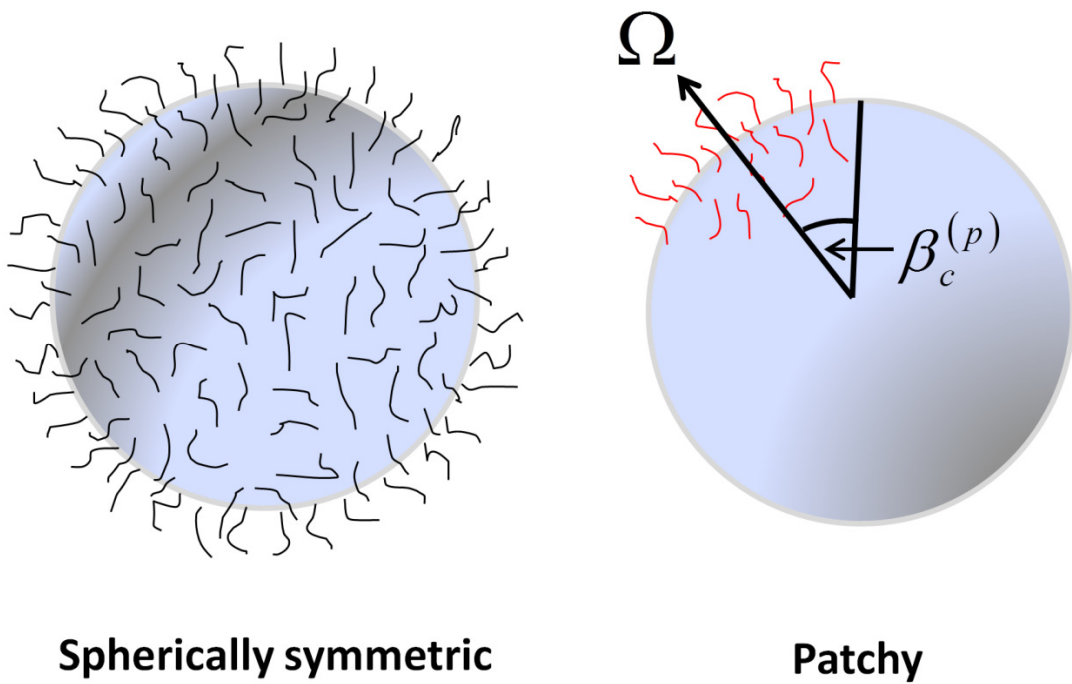

**Figure 1:** Diagram of spherical and patchy colloid. The angle $\beta_c^{(p)}$ defines the size of the patch and $\Omega$ defines the orientation of the patchy colloid. In this depiction we are only illustrating the tethered DNA with sticky ends.

$$\Delta c^{(o)} = \underset{\Delta c_1^{(o)}}{\text{[figure]}} + \underset{\Delta c_2^{(o)}}{\text{[figure]}} + \underset{\Delta c_3^{(o)}}{\text{[figure]}} + \cdots + \Delta c_{n^{\max}}^{(o)}$$

**Figure 2:** Diagram of contributions to the graph sum $\Delta c^{(o)}$, Eq. (4)

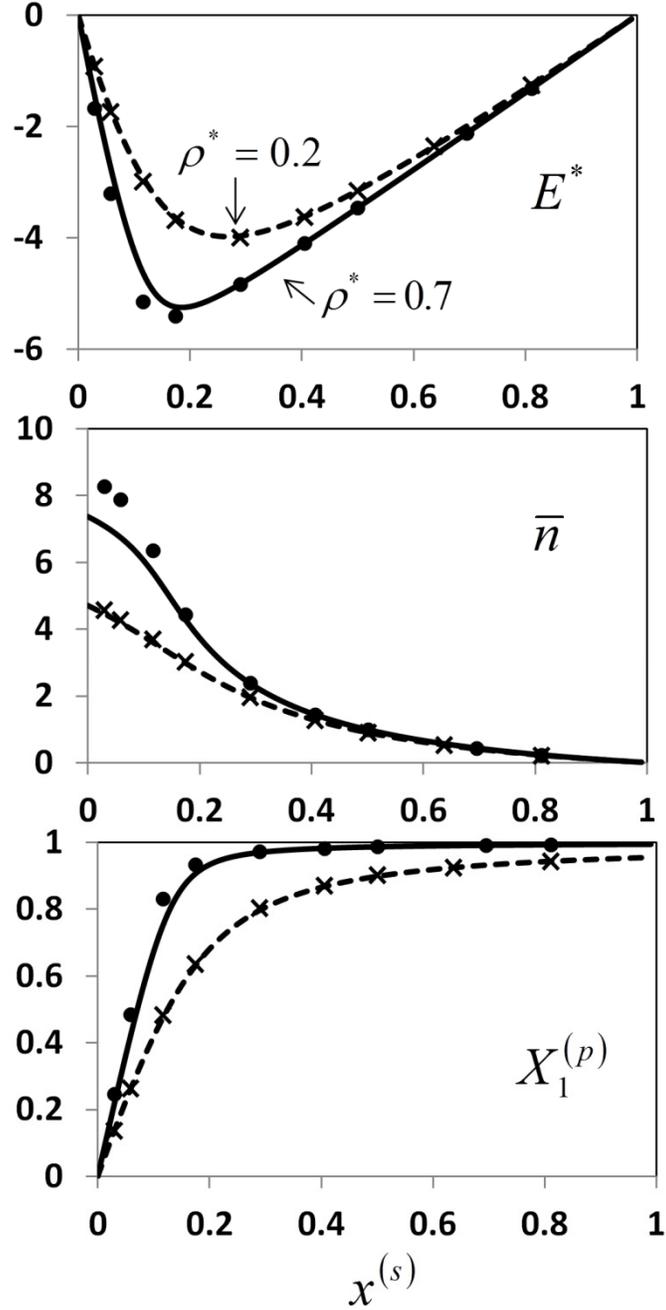

**Figure 3:** Excess internal energy $E^* = E^{AS}/Nk_BT$ (top), average number of bonds per $s$ colloid $\bar{n}$ (middle) and fraction of patchy colloids bonded $X_1^{(p)} = 1 - X_o^{(p)}$ (bottom) versus mole fraction of $s$ type colloids $x^{(s)}$ at an association energy $\varepsilon^* = 7$ and densities $\rho^* = 0.2$ (dashed curve - theory, crosses - simulation) and $\rho^* = 0.7$ (solid curve – theory, circles - simulation)

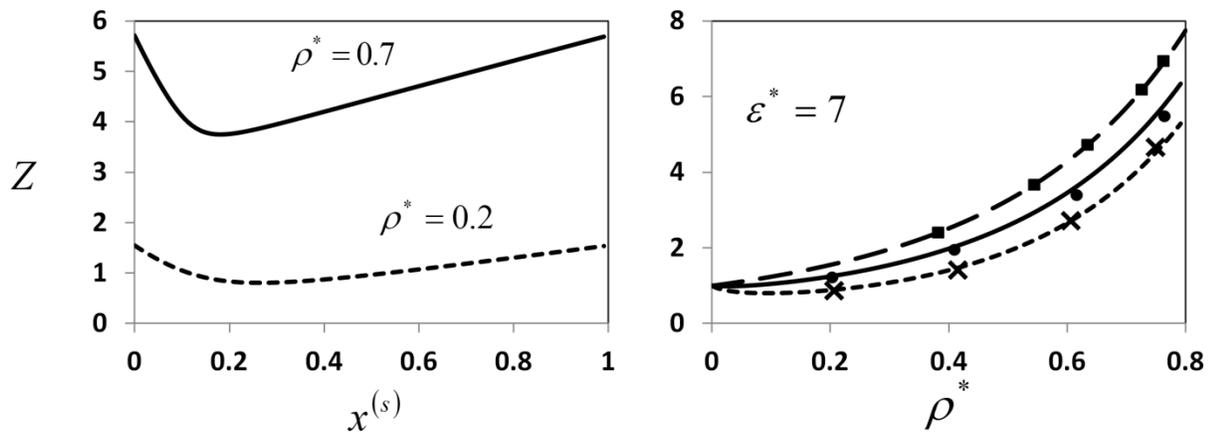

**Figure 4:** Left: Compressibility factor Z versus mole fraction of s type colloids $x^{(s)}$ at an association energy $\varepsilon^* = 7$. Right: Z versus density $\rho^*$ at $\varepsilon^* = 7$ for $x^{(s)} = 0$ (long dashed curve - theory, squares – simulation), $x^{(s)} = 0.0579$ (solid curve - theory, circles - simulation) and $x^{(s)} = 0.174$ (short dashed curve - theory, crosses simulation)

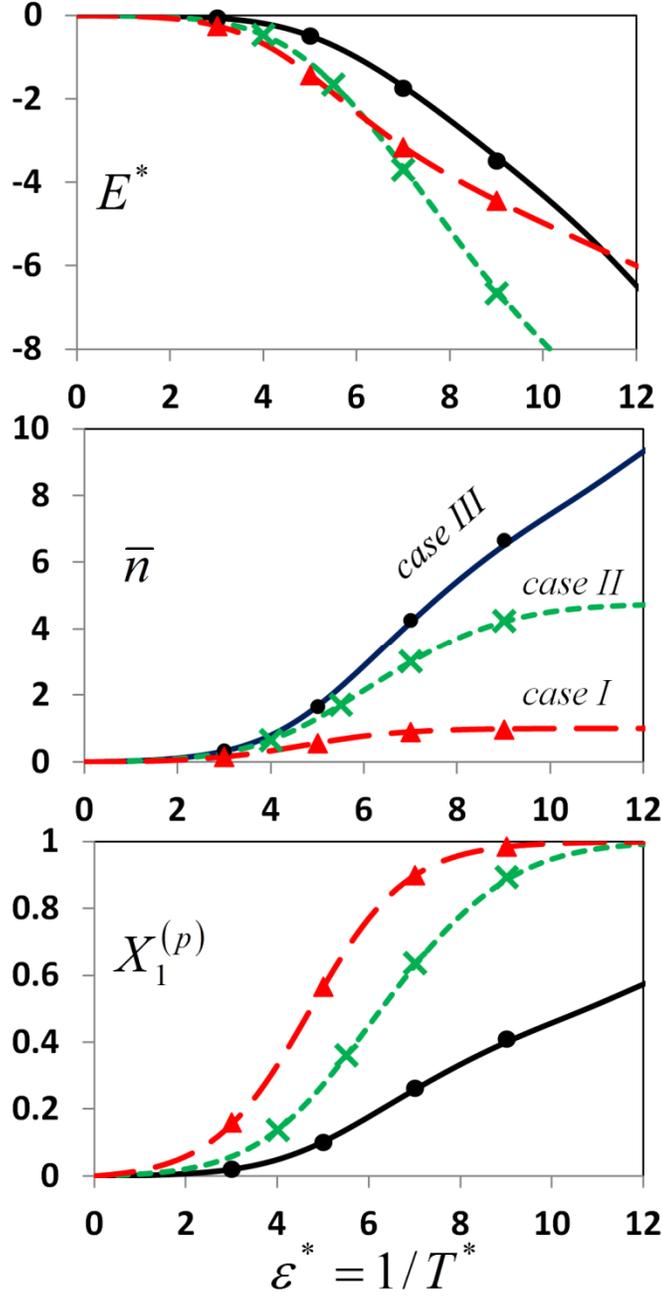

**Figure 5:** Excess internal energy $E^*$ (top), average number of bonds per s colloid $\bar{n}$ (middle) and fraction of p colloids bonded $X_1^{(p)} = 1 - X_o^{(p)}$ (bottom) versus association energy $\varepsilon^*$ at a density $\rho^* = 0.2$ and $x^{(s)} = 0.0579$ (solid curve - theory, circles - simulation), $x^{(s)} = 0.174$ (short dashed curve - theory, crosses - simulation) and $x^{(s)} = 0.5$ (long dashed curve - theory, triangles - simulation)

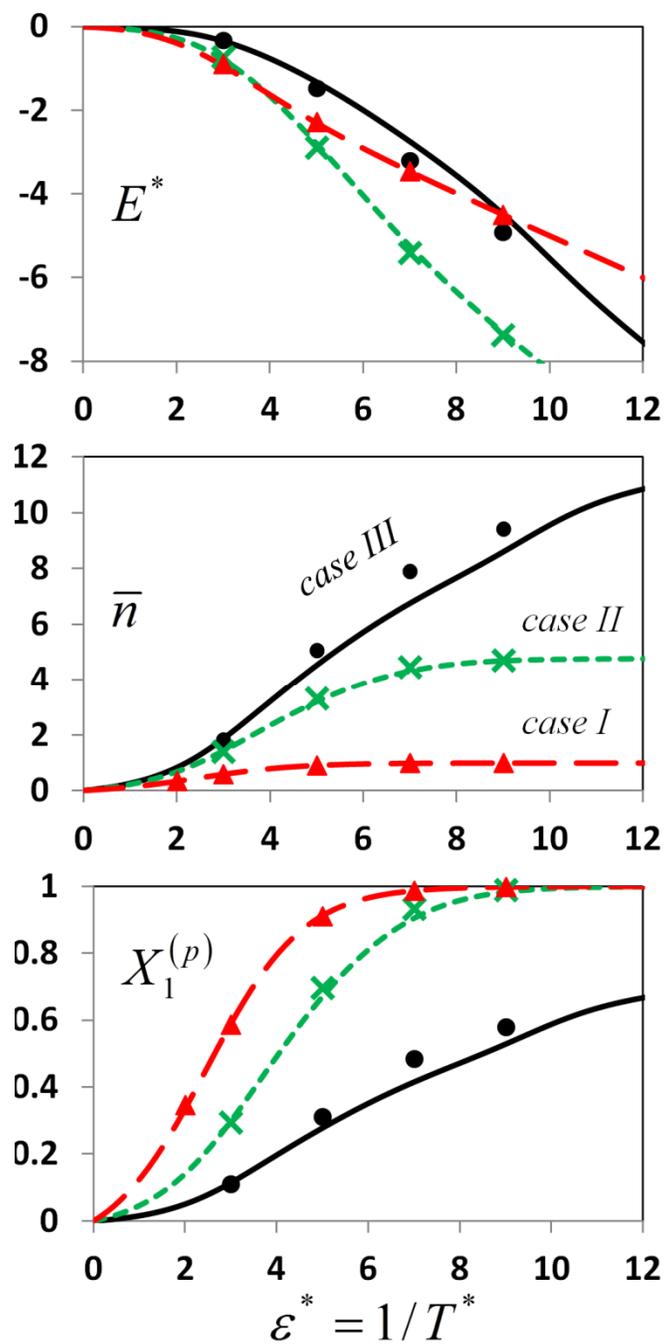

**Figure 6:** Same as Fig. 5 with $\rho^* = 0.7$

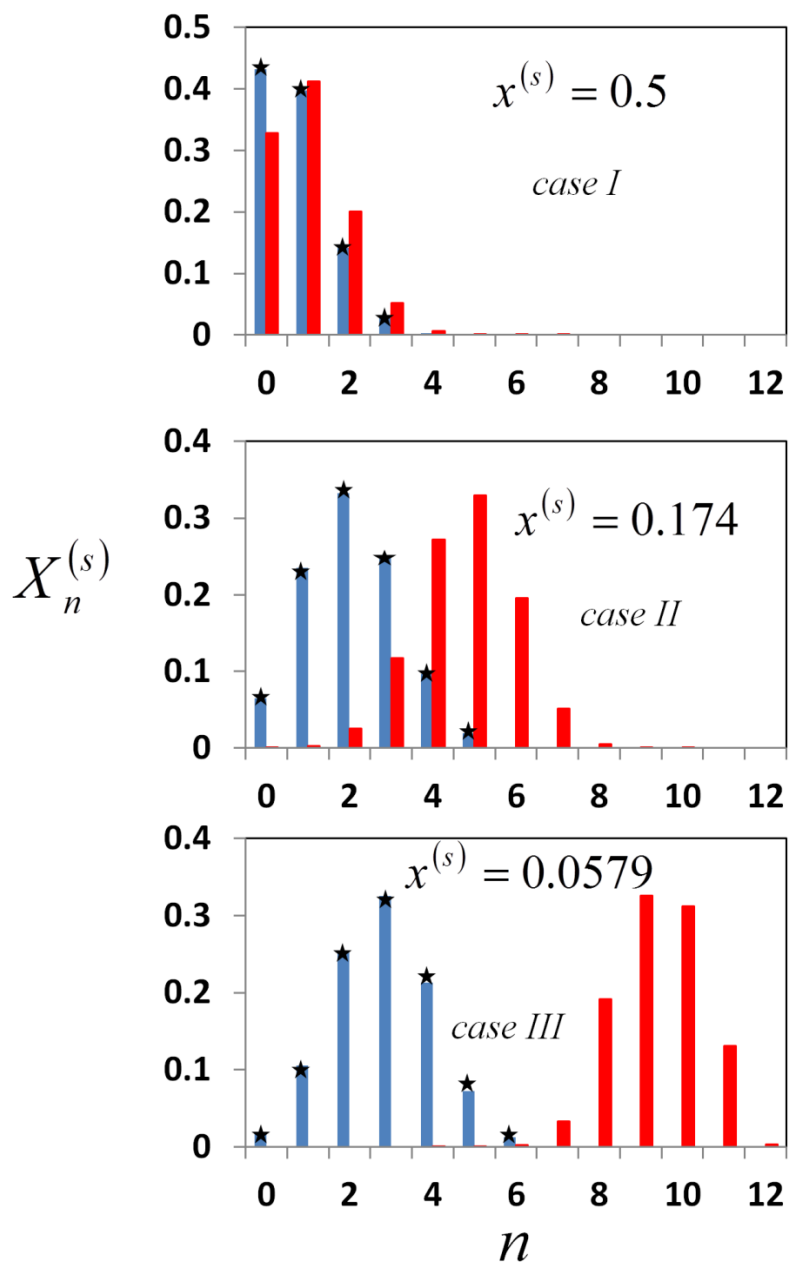

**Figure 7:** Fractions of s colloids bonded $n$ times $X_n^{(s)}$ versus $n$ at a density of $\rho^* = 0.2$ and association energies $\varepsilon^* = 6$ (blue bars – theory, stars – simulation) and $\varepsilon^* = 12$ (red bars – theory)

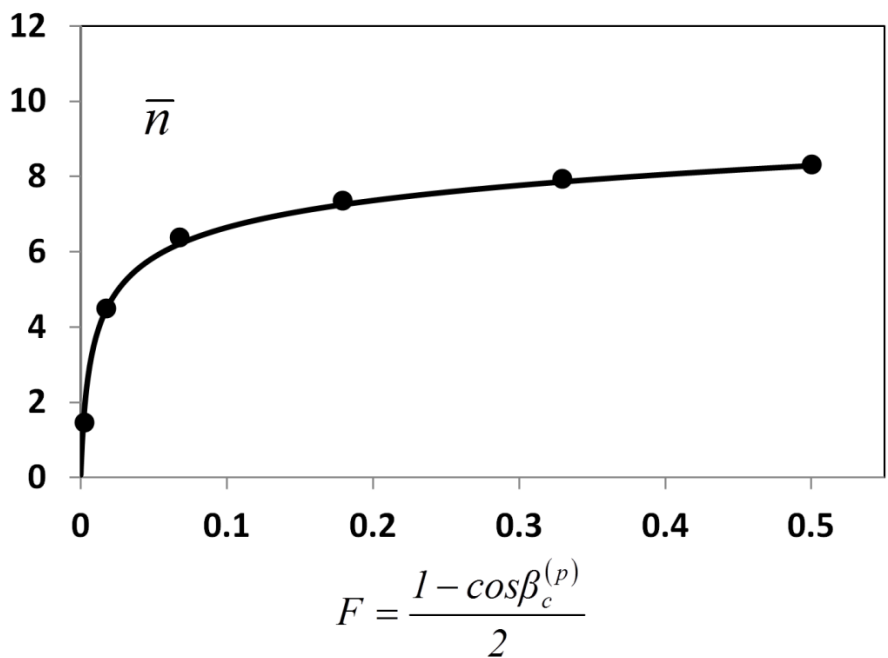

**Figure 8:** Average number of bonds per $s$ colloid $\bar{n}$ versus fractional patch coverage $F$ at a density $\rho^* = 0.2$ and mole fraction $x^{(s)} = 0.00579$. Curve gives theory predictions and symbols are simulation results